# Review of Photoacoustic imaging plus X

**Daohuai Jiang,[a,b,e] Luyao Zhu,[a,e] Shangqing Tong,[a,e] Yuting Shen,[a,e] Feng Gao,[a] Fei Gao[a,c,d,*]**

[a] *ShanghaiTech University, School of Information Science and Technology, Shanghai, China*

[b] *Fujian Normal University, College of Photonic and Electronic Engineering, Fuzhou, China*

[c] *Shanghai Engineering Research Center of Energy Efficient and Custom AI IC, Shanghai, China*

[d] *Shanghai Clinical Research and Trial Center, Shanghai, China*

[e] *Equal contribution*

**Abstract**

Photoacoustic imaging (PAI) is a novel modality in biomedical imaging technology that combines the rich optical contrast with the deep penetration of ultrasound. To date, PAI technology has found applications in various biomedical fields. In this review, we present an overview of the emerging research frontiers on PAI plus other advanced technologies, named as "PAI plus X", which includes but not limited to: PAI plus treatment, PAI plus new circuits design, PAI plus accurate positioning system, PAI plus fast scanning systems, PAI plus novel ultrasound sensors, PAI plus advanced laser sources, PAI plus deep learning, and PAI plus other imaging modalities. We will discuss each technology's current state, technical advantages, and prospects for application, reported mostly in recent three years. Lastly, we discuss and summarize the challenges and potential future work in PAI plus X area.

**Keywords**: photoacoustic imaging, treatment, circuit design, ultrasound sensor, laser source, deep learning, multimodal imaging.

\* gaofei@shanghaitech.edu.cn

## 1 Introduction

Photoacoustic imaging (PAI) is a burgeoning modality within the realm of biomedical imaging, harnessing the dual benefits of rich optical contrast and deep ultrasound penetration[1]. PAI relies on the photoacoustic (PA) effect, facilitating the discernment of signals from within deep tissue layers. The resultant PA signal manifests as an acoustic wave, generated through the illumination of light pulses, typically with nanosecond durations. Over the past two decades, multiple different modalities of PAI systems have undergone rapid advancement, exemplified by PA tomography (PAT), PA microscopy (PAM), and PA endoscopy (PAE)[2], basically. Owing to its multi-wavelength optical absorption contrast, PAI effectively detects both intrinsic and extrinsic optical absorbers, thus cementing its position as a powerful tool for molecular and functional imaging[3].

From technical perspective, PAI is being empowered by, or empowering, other advanced technologies, to achieve lower cost, deeper penetration, faster speed, treatment monitoring, intelligent diagnostics, and more. We name such innovation strategy as: PAI plus X.

In this review, we will summarize recent progress on several important aspects of PAI plus X: 1). PAI plus treatment, such as PAI-guided laser/ultrasound/RF therapy, and surgery, etc. 2). PAI plus advance electrical and mechanical hardware, such as analog/digital circuit, electromagnetic/optical positioning system, mechanical scanning system, etc. 3). PAI plus advanced laser sources, such as adaptive/diffractive optics enabled needle-shape beam, low-cost compact laser diode (LD) and light-emitting diode (LED), high-repetition rate laser sources, etc. 4). PAI plus advanced ultrasound sensors, such as expanding bandwidth by new materials, flexible substrate for wearable sensors, miniaturized CMUT/PMUT sensors, noncontact light-based acoustic sensors, etc. 5). PAI plus deep learning, such as for image reconstruction and enhancement, motion correction and denoising, image analysis and quantification, etc. 6). PAI plus other imaging modalities, including multimodal PA/ultrasound/MRI/Optical imaging, PA-generated ultrasound imaging, ultrasound-assisted PA image reconstruction, etc. Overall, we aim to provide a comprehensive overview of recent advancements in the field of PAI plus X.

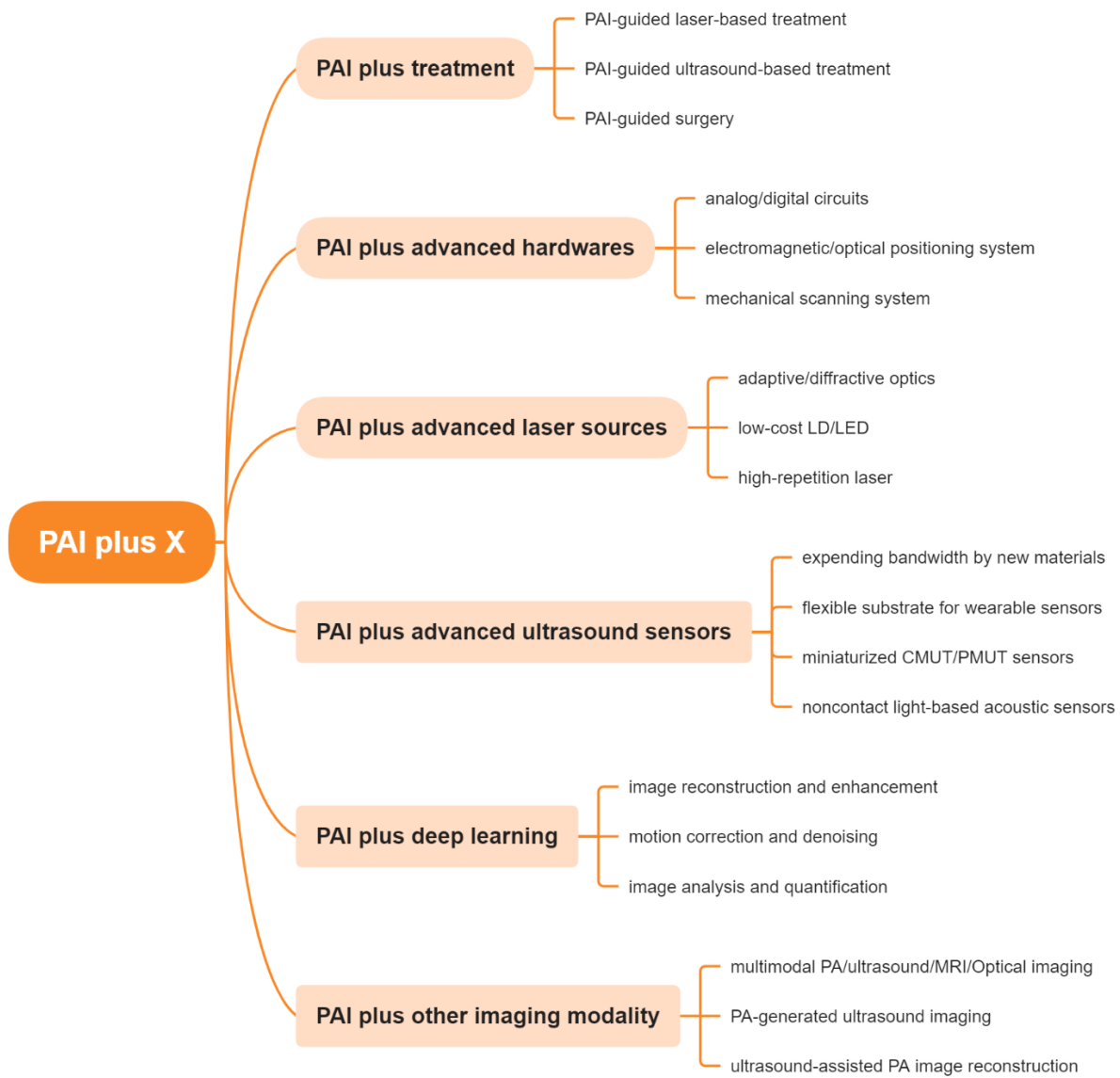

**Fig. 1** Overview of PAI plus X.

## 2 PAI plus Treatment

### 2.1 PAI-guided ablation

Advancements in ablation methods, like laser and radiofrequency ablation, have improved treatments for neural oncology, cardiac arrhythmias, and varicose veins. Ablation offers minimally invasive therapeutic options for conditions such as cancer and arrhythmias. However, immediate procedural assessment and real-time feedback still remains lacking. PAI is emerging as a solution, potentially providing real-time insights into ablation-induced necrosis dimensions and temperature changes during ablation procedures.

Mohammad et al. [4] proposed an integrated PAI-guided laser ablation intracardiac theranostic system that provides real-time tissue ablation, lesion monitoring, and tissue distinguishing capabilities. The system as shown in Fig. 2(a) which offers a low-cost and safer approach for potentially minimizing complications and enhancing treatment procedures. M. Sun et al.[5, 6] developed a multi-wavelength PA temperature feedback photothermal therapy (PTT) system for accurate and safe tumor treatment. Real-time temperature control within the target area achieves 0.56 °C and 0.68 °C accuracy, highlighting its strong application potential. D. Silva et al[7]. presented a multiphysical numerical study of a photothermal therapy performed on a numerical phantom of a mouse head containing a glioblastoma by photoacoustic temperature monitoring. S. Yang et al. [8] developed a non-invasive and high-resolution imaging tool called wavelength-switchable PAM to guide photothermal therapy by mapping tumor microvasculature and nanoparticle accumulation. PAM visualizes tumor microvasculature, guiding PTT implementation and efficacy evaluation.

To distinguish ablated tissue from non-ablated tissue based on their spectrum difference, H.K.Zhang et al.[9] use PAI for real-time visual feedback on tissue ablation. They distinguish ablated from non-ablated tissue through spectral differences, mapping ablation extent and lesion distribution growth (as shown in Fig. 2(c)). S.Beck et al. [10] investigated the safety of using PAI for liver surgeries with a 750 nm laser wavelength and approximately 30 mJ laser energy, and proposes a new protocol for studying laser-related liver damage.

For the ultrasound therapy, L.xiang et al.[11] suggested a hybrid method for treating port wine birthmarks using PAI-guided ultrasound as shown in Fig.2(b). This combines both modalities to target deeper capillaries while minimizing adjacent tissue heating. An array-based high-intensity focused ultrasound (HIFU) therapy system (Fig. 2(d)) integrated with real-time ultrasound and PAI is developed by L. Wang et al[12]. The system can accurately target the treatment spot, flexibly deliver or fast-move focus points in the treatment region, and monitor therapy progress in real-time using PA/US dual-modal imaging.

## 2.2 PAI-enabled ultrasound treatment

Neuromodulation is important for understanding the nervous system and treating neurological and psychiatric disorders. Different techniques like deep brain stimulation, transcranial magnetic stimulation, and electrical stimulation have been used for various conditions. The PA technique uses pulsed laser to generate high-intensity ultrashort ultrasound pulses, allowing for high-resolution imaging of biological structures and potential applications in neuromodulation, nanomedicine delivery, ultrasound-encoded optical focus, and large-volume ultrasound tomography.

H. Yu et al. proposed several kinds of PA transmitter to generate acoustic wave such as dynamic acoustic focusing PA transmitter[13, 14] (as shown in Fig. 2(e)), flat PA source based ultrasound transmitter[15], and based on binary amplitude switch control of PA transducer toward dynamic spatial acoustic field modulation[16]. M. Jezerresk et al. proposed an ultrasonic PA emitter by a graphene-nanocomposites film on a flexible substrate[17]. The proposed strategies provide effective methods for dynamically manipulating the acoustic field in PA transmitters, which can have significant applications in various fields. Z. Du et al. presented the development of a candle soot fiber optoacoustic emitter (CSFOE) that can generate high pressure of over 10 MPa with a central frequency of 12.8 MHz, enabling highly efficient neuromodulation in vitro[18]. The CSFOE can perform dual-site optoacoustic activation of neurons, confirmed by calcium ($Ca^{2+}$) imaging, and opens potential avenues for more complex and programmed control in neural circuits using a simple design for multisite neuromodulation in vivo. Precise drug delivery is important for internal organs, L. Xi et al.[19] developed a dual-wavelength photoacoustic laparoscope for nanomedicine delivery, and it shows the optical-resolution PAM (OR-PAM) based precise drug delivery method is promising for the effective treatment of internal organ diseases, Fig. 2 (f) shows the imaging system schematic. Optical imaging is limited by scattering and has a great challenge for deep tissue imaging, J. Zhang et al. proposed wavefront shaping method based on time-reversed ultrasonically encoded optical focusing by PA wave[20], achieving dynamic focusing of light into both optical and acoustic heterogeneous scattering medium, which shows high potential for transcranial light focusing. PA wave has broad bandwidth, and it can be used for ultrasound imaging potentially. S. Manohar et al. proposed laser-induced ultrasound transmitters (LIUS) for large-volume ultrasound tomography[21], The LIUS transmitters produced a center frequency of 0.94 MHz with a bandwidth from 0.17 to 2.05 MHz, producing pressures between 180.17 kPa and 24.35 kPa for a range of depths between 7.42 and 62.25 mm.

*2.3 PAI-guided surgery*

Surgery requires multimodal medical imaging information to improve the efficiency and success rate, which inspires the motivation to develop PAI-guided surgery. These endeavors encompass diverse areas, including PAI-guided surgical procedure, PAI-guided accurate biopsies and tissue characterization.

L. Zhu et al [22] introduced the application of PAI for surgical navigation in spinal surgery procedures. Through a combination of theoretical analysis and experimental verification, the authors demonstrated the feasibility of this approach. For the real-time surgical guidance, E. M. Boctor et al.[23] presented a real-time intraoperative surgical guidance system employing PA markers. This system co-registers a Da Vinci surgical robot's endoscope camera with a transrectal ultrasound (TRUS) transducer (as shown in Fig. 2 (g)). It enables functional guidance within the surgical region-of-interest (ROI) by tracking the pulsed-laser-diode-illuminated laser spot on the surgical instrument.

Combined with PAI to distinguish between tumor and normal tissue, Y. Shi et al. [24] proposed a 532/266 nm dual-wavelength PAM imaging system that can simultaneously perform in vivo analysis of peritumoral vasculature and ex vivo surgical margin pathology of tumors. The system has the potential to guide the process of tumor resection, improve the efficiency of complete tumor resection in a single surgery, and reduce the recurrence rate. V.V Verkhusha et al.[25] presented a transgenic mouse model with a knocked-in BphP1 soluble bacterial NIR photoreceptor. The mouse model enables both spatiotemporal optogenetic regulation and PAI in deep tissues using the same genomically integrated BphP1 construct. The study validates the optogenetic performance of endogenous BphP1 and demonstrates PAI's capability of BphP1 expression in different organs, developing embryos, virus-infected tissues, and regenerating livers. W. Xia et al.[26] developed a compact, high-speed PA endomicroscopy probe capable of real-time visualization of tissue's functional, molecular, and microstructural features. Integrated into a medical needle cannula, this probe shows promise in guiding minimally invasive procedures like tumor biopsies. They also proposed a deep learning framework based on U-Net to improve the visibility of clinical metallic needles with a LED-based PA and ultrasound imaging system.[27] The application of photoacoustic technology in surgical contexts is both imperative and holds significant potential for future developments.

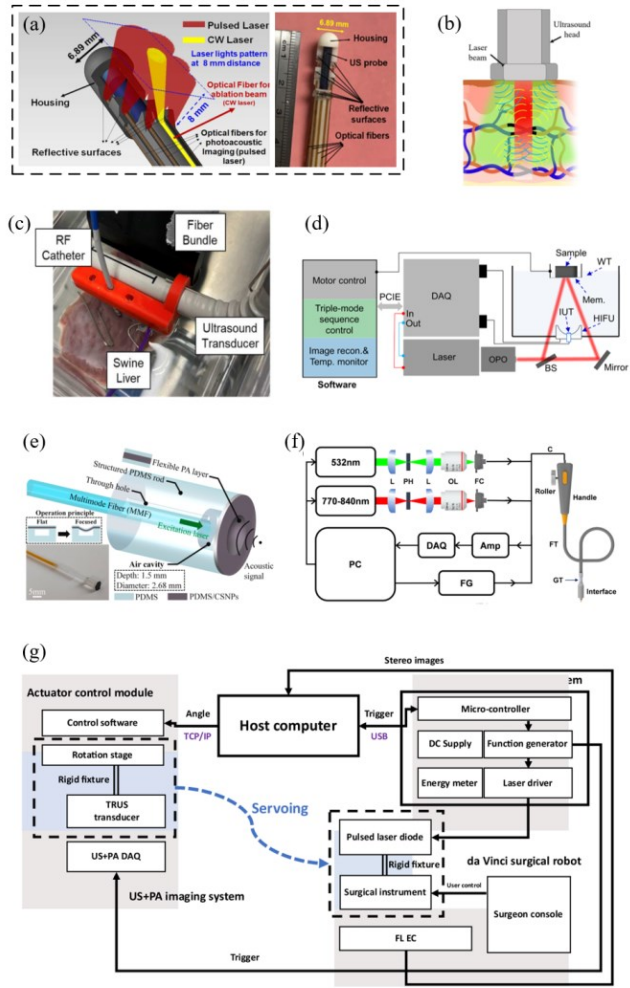

**Fig. 2** The innovation of PAI plus treatment (a) A schematic And photograph of miniaturized integrated US/ PA-guided laser ablation theranostic system[4]. (b)PA guided US therapy with optimal benefits[11]. Yellow wavefronts: PA waves sensed by the transducers; Blue wavefronts: US transmitted by the transducers. PA, photoacoustic; US, Ultrasound. (c)the radiofrequency ablation monitoring with photoacoustic necrotic region mapping[9]. (d) schematic of the tri-modal system for HIFU therapy[12]. (e) schematic of PA transmitter probe by H. Yu et al[14]. (f) the system setup of the OR-PAM system for nanomedicine delivery proposed by L. Xi et al.[19] (g) The overall system architecture of TRUS+PA image-guided surgical guidance system.[23]

## 3 PAI plus Advanced Electrical and Mechanical Hardware

Achieving high-quality PA images necessitates capturing PA signals effectively, which involves various key hardware components, such as analog and digital circuits for signal amplification, denoising, and acquisition. Additionally, ultrasound probe's positioning is crucial for accurate image registration, and the integration of advanced mechanical scanners can significantly expedite the imaging process. In this section, we will delve into the current research landscape concerning PAI plus advanced hardware, encompassing analog/digital circuits, accurate positioning systems, and innovative scanning mechanisms.

### 3.1 Analog and digital circuits enabled low-cost PAI

Photoacoustic signals' capture mainly includes acoustic sensing, amplification, and data acquisition. To achieve higher detection sensitivity with miniaturized size, Y. Zheng et al. presented two cutting-edge approaches in coherent PA sensing technology. The first work[28] involves a silicon-based sensing system-on-chip (SoC) designed for precise in vivo sensing and imaging, particularly for deep vessel and blood temperature assessments. This SoC, made using TSMC 65-nm CMOS technology, holds promise for healthcare monitoring and early disease detection. The second work[29] introduced a quadrature adaptive coherent lock-in (QuACL) chip, a compact chip-based PA sensor utilizing adaptive coherent lock-in techniques for accurate PA signal detection in challenging conditions (as shown in Fig. 3(a)). While its implementation requires two analog-digital conversion boards, it has the potential to be integrated into wearable healthcare devices in the future.

There are also many attempts to reduce the system cost, X. Ji et al[30] proposed a low-cost and compact PA maximum amplitude projection (MAP) microscopy system based on a custom-made peak holding circuit (as shown in Fig. 3(b)), which allows for ultra-low data sampling. The system has the same imaging ability as conventional PAM systems. It provides a new paradigm for PAM and offers a cost-effective solution for optimal PA sensing and imaging devices. In pursuit of high-fidelity PA images, capturing an increased number of PA signals escalates hardware costs. To mitigate data acquisition (DAQ) channel consumption, D. Jiang et al[31-34] introduced various time and frequency division multiplexing approaches. Examples include the multi-channel delay line module and the low-cost PAT system based on frequency division multiplexing. These methods effectively curtail the DAQ system's cost.

To reduce the cost of computation of PA image reconstruction, Y. Shen et al[35] introduced a faster model-based image reconstruction method based on superposed Wave (s-Wave). The proposed method demonstrates substantial time savings, particularly in sparse 3D configurations, where it is over 2000 times faster. To realize fast and cost-effective image reconstruction of PAI. Z. Gao et al[36] proposed a palm-size and affordable PAT system[36] with hardware acceleration. The system employs field-programmable gate array (FPGA) implementation for high-quality image reconstruction in low-cost low-power FPGA platform, which is adaptable to various image reconstruction algorithms, accelerating the reconstruction speed at the much less system cost.

*3.2 Positioning system enhanced PAI*

The reconstructed PA image's quality relies on the accurate positioning of the ultrasound probe. To ensure image alignment and accurately register the images with the imaging target, various techniques have been employed. Notably, for handheld PAI systems, approaches like optical camera-based positioning and electromagnetic field-based global positioning system (GPS) have been employed to tackle this challenge. L. D. Liao et al[37] introduced a new PAI system called ViCPAI that combines a visible CCD camera with an ultrasound transducer for precise positioning and imaging in preclinical and clinical studies. The system accurately locates target areas and achieves reproducible positioning, allowing for real-time capturing of cerebral hemodynamic changes during various experiments, such as forelimb stimulation and stroke induction. It also enables the monitoring of cortical spreading depression (CSD) and the progression of peri-infarct depolarization (PID) after stroke. The ViCPAI system overcomes the limitations of existing imaging systems by providing precise positioning capabilities and an intuitive user interface.

D. Jiang et al[38, 39] proposed a handheld free-scan 3D PAT system (fsPAT) for clinical applications. Using a linear-array ultrasound probe coordinated via electromagnetic field-based GPS systems, it achieves real-time 2D imaging and large field-of-view 3D volumetric imaging (as shown in Fig. 3(c)). A specialized space transformation method and reconstruction algorithm could enhance 3D image quality. In vivo human wrist vessel imaging demonstrates the clinical potential of fsPAT, revealing detailed subcutaneous vessels with high image contrast. For the PAM system, L. Wang et al.[40] introduced FS-PAM, a handheld probe overcoming traditional optical-resolution photoacoustic microscopy limitations like limited field of view, bulky probes, and slow speed. FS-PAM uses a hybrid resonant-galvo scanner for high-speed dual-axis scanning, offering high-resolution, motion artifact-reduced, label-free hemodynamic and functional imaging. Real-time imaging and simultaneous localization and mapping (SLAM) mode are possible due to its high scanning speed. FS-PAM's success is exemplified in imaging mouse organs, human oral mucosa, localizing brain lesions and stroke models.

To precisely align imaging results with specific imaging targets, such as blood vessels, S. Yang et al. [41] introduced a method that combines PAT and optical projection for noninvasive high-resolution imaging of deep blood vessels in the human body. By aligning PA data with real patient anatomy, this technique enables three-dimensional visualization of blood vessels from the body surface. The system has guided micro plastic injection and revealed submillimeter forehead blood vessels, showing potential for aesthetic medicine.

### 3.3  Advanced scanning mechanisms

PAI relies on ultrasound transducers to receive PA signals from various positions. Notably, in PAM systems, the need to raster-scan all imaging pixels often results in time-consuming data acquisition processes. Consequently, employing advanced scanning mechanisms can substantially enhance imaging speed. C. Liu et al.[42] presented an improved multiscale PAM system that achieves high-speed wide-field imaging using a homemade polygon scanner (as shown in Fig. 3 (d)). The system overcomes the tradeoff between imaging speed and field of view in previous PAM systems and demonstrates increased imaging speed by a factor of 12.35 compared to previous systems. Y. Saijo et al. [43] introduced a novel and simple distortion correction method for high-speed OR-PAM with micro electromechanical system (MEMS) scanner. J. Yao et al. [44] introduced a high-speed functional photoacoustic microscopy (OR-PAM) system employing a water-immersible two-axis torsion-bending scanner (as shown in Fig. 3(e)). This innovation accelerates traditional OR-PAM imaging by enabling rapid 2D scanning with independent adjustments of scanning speed and range along both axes. With a cross-sectional frame rate of 400 Hz and volumetric imaging speed of 1 Hz across a $1.5 \times 2.5$ mm$^2$ field of view, the system effectively captures dynamic information in small animal models in vivo, including hemodynamic changes under pharmaceutical and physiological influences.

For more dedicated applications of PAM, L. Xi et al. [45-47] have introduced several interesting PAM platforms with versatile applications. The Organ-PAM platform enables high-resolution imaging of multiple vessel systems within organs, revealing insights into pathological conditions.

An ultrafast functional photoacoustic microscopy (UFF-PAM) system achieves real-time whole-brain imaging of hemodynamics and oxygenation at micro-vessel resolution, showing potential for fundamental brain research[46]. Additionally, they present a detachable head-mounted PAM probe for optical-resolution imaging in freely moving mice[47], offering stable imaging of cerebral dynamics. These platforms collectively push the boundaries of PAI, offering a spectrum of capabilities from organ-level visualization to dynamic brain studies.

For multi-wavelength PAM, M. Ishihara et al.[48] proposed a new spectroscopic OR-PAM technique that allows for the acquisition of information on PA signal intensity and excitation wavelength from a single spatial scan. The technique involves using two broadband optical pulses with and without wavelength-dependent time delays to calculate the excitation wavelength of the sample. The combination of this technique with fast spatial scanning methods can significantly contribute to recent OR-PAM applications. The limited detection view and detection bandwidth in OR-PAM can lead to a nonlinear dependence on optical absorption, especially for weakly absorbing targets. The study by T. Duan et al.[49] proposed potential solutions and confirms the results through numerical simulations and experimental validation on phantoms. The findings suggest that a side detection scheme or a high optical numerical aperture (NA) may mitigate the low detection sensitivity of OR-PAM on weakly absorbing targets

K. Xiong et al.[50] presented the development of a flexible forward-view photoacoustic endoscopy (PAE) probe based on a resonant fiber scanner, which enables noninvasive biopsy in narrow areas of internal organs. The probe integrates a piezoelectric bender, a fiber cantilever, a lens, an ultrasound transducer, and a coupler, and achieves a lateral resolution of 15.6 μm in a field-of-view of approximately 3 mm diameter. Its potential lies in serving as a minimally invasive tool for the clinical assessment of gastrointestinal lesions.

Y. Zhao[51] proposed a novel design for a continuously-adjustable light-scanning handheld probe for PAI, which can acquire multiple images using different illumination schemes, which can be easily held with one hand. The probe consists of three parts: a medical US linear-probe clamp, a light transition unit, and an optic wedge unit for light beam shaping. The design allows for adjusting the illumination schemes according to different samples, addressing the issue of delivering more photons to deeper tissues without exceeding safety standards or causing overexposure.

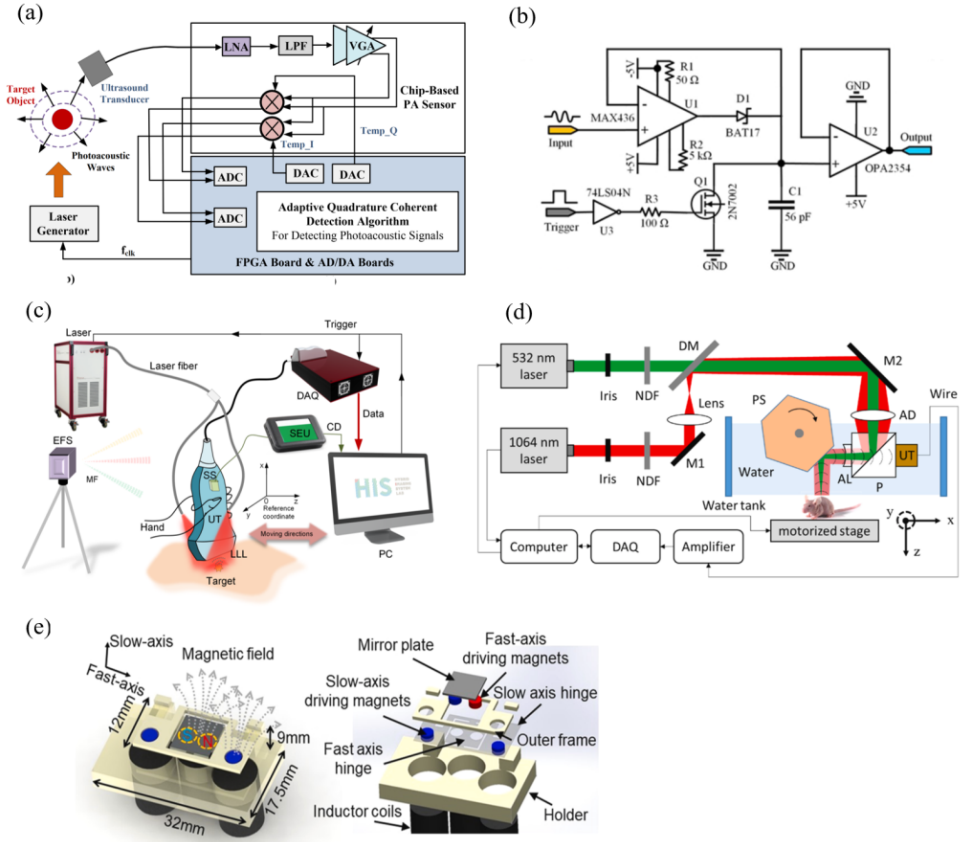

**Fig. 3** The innovation of PAI plus electrical and mechanical hardware (a) system diagram of QuACL chip-based PA detection.[29] (b) schematic diagram of the peak holding circuit.[30] (c) The setup of the 3D fsPAT imaging system.[38] (d) schematic of the multiscale PAM with polygon-scanning method.[42] (e). working principle of the torsional-bending scanner mentioned in Ref [44].

## 4 PA Plus Advanced Laser Sources

In PAI, the target tissue absorbs thermal energy under the illumination of a laser source, leading to energy conversion and the generation of PA signals that carry tissue's molecular information. In this process, the selection of the laser source directly determines the efficiency and quality of signal generation, and has a significant impact on the overall system cost. Throughout the development of PAI, researchers have continuously explored and improved the selection of laser sources, considering factors such as laser source types, size, costs, repetition rate, and so on, to enhance PAI's performance and expand potential applications.

*4.1 Enhancing Resolution*

Multiple techniques are presented for overcoming limiting factors in spatial resolution to improve visualization of fine structures and details. The ultimate resolution achieved in PAI is dependent on both the optical excitation and acoustic detection. Therefore, many researches have been focusing on the improvement of resolution by innovating laser sources. Nteroli et al. and NOTSUKA et al. improve axial resolution through novel laser sources [52] and adaptive optics [53], respectively. Nteroli et al. shows ultrafast picosecond excitation can generate ultrasound waves of higher frequency for up to 50% improvement in axial resolution. Adaptive optics [53] helps compensate aberrations to maintain tight optical focus and high lateral resolution at depth for OR-PAM (as shown in Fig. 4 (a)). Cao et al. develops needle-shaped optical excitation beam using diffractive optical elements to extend depth of field to around 28-fold Rayleigh length [54]. This helps enable high resolution imaging over an extended imaging volume without the need for fine depth scanning and focus adjustment.

These advancements push the boundaries of the axial resolution possible and demonstrate the potential for PAI to achieve diffraction-limited resolution sufficient for resolving subcellular structures, even in deep tissues. It allows maintaining tight lateral resolution when imaging uneven surfaces or acquiring volumetric data.

*4.2 Shrinking Cost and Size*

It is important for a PAI system to be both competitive in imaging quality and accessible for more patients. Therefore, lower system costs combined with the smaller size of laser sources facilitates development of portable imaging systems that can be moved close to the patient bedside or surgical suite. Li et al.[55] demonstrates a compact dual-laser diode PAM with a cost reduction of nearly 20-40 times compared to standard pulsed laser systems typically used for PA excitation (as shown in Fig. 4 (b)). The laser diodes provide sufficient pulse energy for high resolution OR-PAM. Replacing costly and bulky pulsed lasers with inexpensive compact laser diodes could help expand adoption of PAI. Similarly, Heumen et al. explores the use of low-cost light-emitting diodes (LEDs) as excitation sources for visualizing lymphatic vessels in patients with secondary limb lymphedema [56]. Deng et al. [57] and Liang et al. [58] also showcase portable, low-cost laser diode and LED-based photoacoustic imaging systems tailored for specific clinical applications like subsurface microvasculature or lymphatic imaging. Deng et al. develops a laser diode based system with a long working distance of 22 mm using reflective optics[57]. The long working distance enables non-contact imaging and overcomes limitations of the short working distances of high numerical aperture objectives needed to focus laser diode beams. Moreover, Song et al.[59] develops a multiscale technique that can tune resolution by controlling the spatial frequency of structured illumination patterns, avoiding slow mechanical scanning.

The low cost and small size of laser diodes and LEDs makes them well-suited for translating PAI to bedside use. Their efficiency and reduced power requirements also enable development of compact portable imaging systems.

*4.3 Increasing Imaging Speed and Frame Rate*

Boosting PAI's speed enables real-time visualization of dynamic physiological processes across different time scales. Chen et al. achieves video-rate 30 Hz PAI over a 473 μm field of view by using non-scanning single pixel detection[60]. This represents nearly two orders of magnitude increase in speed over conventional scanning OR-PAM. Real-time PAI could enable new research and clinical capabilities ranging from imaging blood flow to tracking cancer cell metastases. Such speed improvement is realized by combining single-pixel detection with customized temporally modulated illumination patterns, overcoming limitations imposed by conventional raster scanning. Guo et al. and Song et al. apply computational approaches like compressed sensing[61] (as shown in Fig. 4 (c)) and Fourier basis encoding[59] to accelerate data acquisition and reconstruction to help overcome speed limitations of conventional scanning and reconstruction methods. Compressed sensing can provide high fidelity reconstruction from sparse sampling by exploiting image sparsity and redundancy. Fourier basis encoding utilizes predictable mathematical patterns to enable reconstruction from limited detection data. These computational innovations could be combined with alternative detection schemes to achieve real-time PAI.

*4.4 Enabling More Clinical Applications.*

Advanced laser sources are key to expanding the imaging capabilities and applications of PAI, from preclinical animal models to human patients. Lipid detection could provide critical information about disease progression in conditions like atherosclerosis. Multispectral imaging allows better differentiating lipid, hemoglobin, and other absorbers. Ren et al.[62], Mukhangaliyeva et al.[63] and Liang et al.[58] showcase handheld, non-contact, and other systems tailored for clinical use. The handheld probe in Ref[62] enables detecting optical anisotropy for assessing tissues like nerves and tendons during surgical procedures. Non-contact imaging in Ref[63] (as shown in Fig. 4(d)) and Ref[58] helps prevent contamination and damage to delicate samples. Lee et al. demonstrates a specialized fiber laser providing picosecond pulses at 1192 nm for PAM of lipids in the second near-infrared window[64]. Operating in this optical window allows deeper penetration for lipid imaging. Tachi et al. uses a supercontinuum source for chromatic aberration-free multispectral imaging[65]. Moreover, S.V Heumen et al.[56] demonstrated the promise of translating PAI to patients by visualizing lymphatic vessels in human subjects with LED excitation. As the technology matures gradually, clinical translation from preclinical studies to clinical use for improving patient care and outcomes could accelerate.

In summary, ongoing innovations in laser sources are helping make PAI systems more accessible, faster, higher-resolution, and better suited for clinical translation. With its unique ability to provide high resolution optical absorption contrast deep in tissues, PAI is poised to become a valuable new tool for both biomedical research and clinical diagnosis.

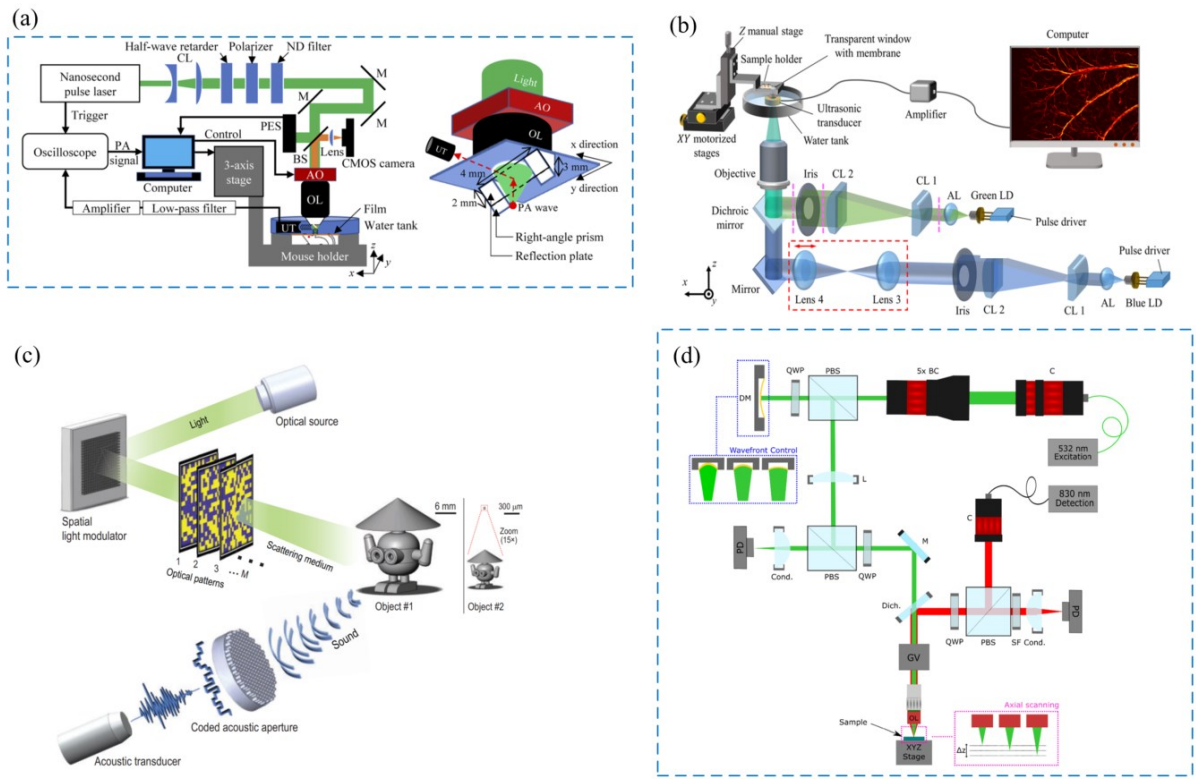

**Fig. 4** Novel laser sources plus PAI. (a) The AO-PAM system and its high-NA objective lens and a narrow reflection plate[53] (b) The system schematic of the dual-wavelength LD-based PAM.[55] (c) A schematic diagram of dual-compressed photoacoustic single-pixel imaging.[61] (d) the simplified photoacoustic remote sensing microscopy optical setup.[63]

## 5 PA plus Advanced Ultrasound Sensors

Typically, a dedicated ultrasound transducer needs to be used for PA signal acquisition, enabling subsequent signal processing and analysis. Due to variations in tissue morphology and composition, the requirements for the ultrasound sensors can differ. Significant differences in tissue size and morphology across different regions necessitate adjustments in the material, size, shape, and arrangement of individual elements within the ultrasound transducer to ensure both detection accuracy and convenience. Recent researches highlight a series of ultrasound sensor innovations, including novel geometries, materials, fabrication methods, and sensing principles.

### 5.1 Sensor Bandwidth and Frequency Range

Due to the different optical and acoustic characters of different types of tissues, the frequency spectrum of PA signals emitted by tissues may exhibit noticeable differences, leading to varying demands on the central frequency and bandwidth of the ultrasound transducer. Several works have pushed the boundaries of sensor bandwidth and frequency range to overcome limitations of conventional piezoelectric detectors (as shown in Fig. 5 (a)). Multi-frequency or broadband detectors are implemented to capture a wider range of PA signal's frequencies[66-69]. Broader bandwidths improve imaging resolution, allow access to higher frequency content, and enable spectroscopic PA analysis. Spectroscopic detection analyzes the acoustic frequency spectrum at each pixel to extract additional information about the optical absorber for enhanced tissue characterization. However, achieving wide, flat sensor bandwidths spanning tens to hundreds of MHz still remains technically challenging.

### 5.2 Fabricating Sensors on Transparent Substrates

Fabricating sensors on transparent substrates is another important theme[70-73]. Transparent sensors enable co-axial or trans-illumination optical delivery, which is critical for some implementations like handheld probes. They also facilitate multimodal imaging, allowing flexible ultrasound detector integration with other optical modalities. This expands possibilities for multimodal imaging, allowing flexible ultrasound detector integration within optical microscopy setups.

However, fabrication of transparent ultrasound sensors with adequate sensitivity and bandwidth has been a persistent challenge. Materials and designs to improve transparent sensor bandwidth, sensitivity, and acoustic matching are investigated. Osman et al.[71] explores using dispersed glass microbeads in epoxy to create acoustic matching layers. Meanwhile, Chen et al.[70] and Peng et al.[72] apply various piezoelectric materials like lithium niobate for making transparent ultrasound sensors and transducers. Realizing transparent detectors with adequate acoustic characteristics would allow flexible integration into multi-modal imaging systems and open up novel implementation methods.

*5.3 Sensor Geometries and Arrangements*

Beyond fundamental materials and fabrication advances, researchers are also innovating sensor geometries and arrangements to enhance PAI. Zhang et al.[74] and Ma et al.[75] exploit non-spherical elliptical and needle-shaped focusing to achieve extended depth of field. Dense sensor arrays and transparent array[76] (as shown in Fig. 5(b)) with up to thousands of elements are also developed, enabling real-time image acquisition without mechanical scanning[69, 76-79]. Fu et al.[80] presents a hockey stick shaped sensor tailored for intraoral imaging of posterior teeth, which are not easily accessed by conventional linear transducer geometries. These works demonstrate how purposefully engineering sensor geometry can improve imaging performance and expand potential applications. Geometry innovations address specific use cases and limitations, demonstrating how customizable, application-specific sensor design can maximize imaging performance.

*5.4 Miniaturized Sensors*

Another major advancement is the development of miniaturized sensors that facilitate integration into compact imaging systems and enable new minimally invasive applications (as shown in Fig. 5(d)). Several studies present miniaturized optical fiber, silicon, and PMUT based sensors with dimensions less than 1 mm, suitable for catheter or endoscopic imaging[81-84]. Such miniaturized sensors overcome limitations of bulky piezoelectric detectors and open new possibilities for invasive imaging. These sensors can be placed at the tip of needles and catheters to provide high resolution in vivo imaging during guided interventions, or implemented in thin endoscopic probes for assessing internal hollow organs.

However, scaling down sensor size can compromise sensitivity and bandwidth if not properly designed carefully. Therefore, researchers are also exploring novel materials and fabrication techniques to maintain adequate acoustic performance in miniaturized footprints. Ustun et al.[82] utilizes high frequency silicon-based acoustic delay lines to achieve 20 MHz bandwidth for its sub-mm fiber optic sensor. Wang et al. [83] and Cai et al. [84] exploit the favorable scaling of PMUT technology to enable micromachined arrays for endoscopic imaging. Ongoing efforts to optimize miniaturized sensor design while preserving sensitivity, bandwidth, and other acoustic characteristics will enable translation to clinically valuable minimally invasive imaging applications.

*5.5 Optical Sensors*

Several works also showcase optical detection schemes that offer advantages over conventional piezoelectric sensors. Although optical sensing modalities can have tradeoffs like anti-interference capability and more complex nanofabrication, which may hinder its broader adoption, improvements have been made to improve optical detector's performance to make them more viable alternatives. Optical interferometry, surface plasmon resonance, and related principles are utilized to achieve acoustic sensing [68, 75, 85-95]. Compared to piezoelectric sensors, optical detection provides higher sensitivity, larger bandwidths extending to hundreds of MHz, smaller footprints without electrical connections, and easier integration with optical excitation sources. Optical sensing is also free of acoustic impedance mismatches, making the signal detection more convenient and efficient.

The growing number of in vivo imaging demonstrations reveal the expanding practical applications as sensor technology matures. Researches in [83, 84, 96] present endoscopic and other minimally invasive sensor configurations tailored for specific clinical use cases and imaging needs. Meanwhile, researchers of [75, 81, 93-95] directly showcase in vivo imaging enabled by optical or miniaturized sensors, including a brain imaging demonstration. Translation to preclinical and clinical imaging is critical to validate the sensors and make real-world impact.

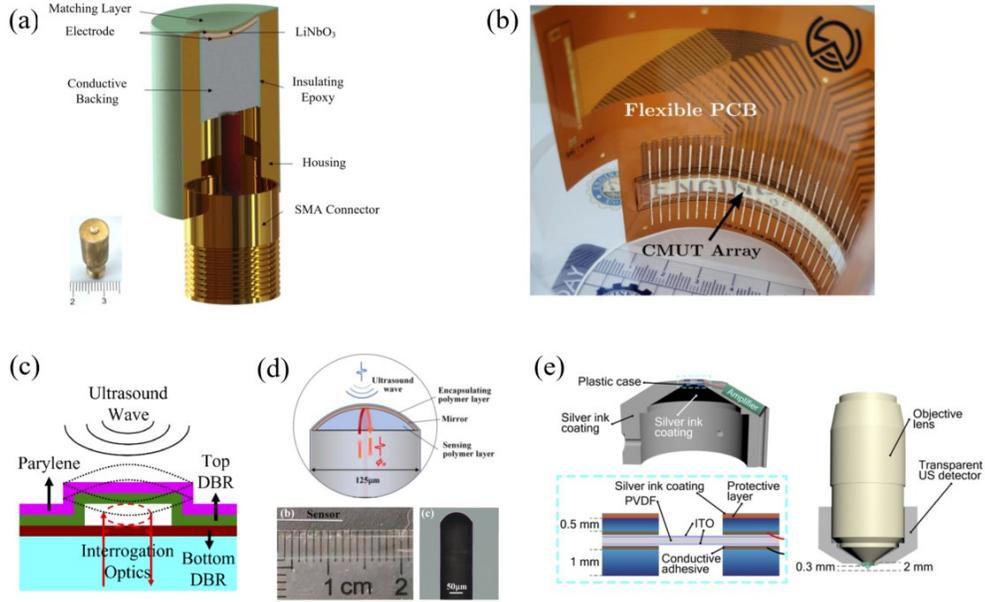

**Fig. 5** Advance ultrasound sensors for PAI. (a) Cross-sectional structure of the broadband transducer and a photograph of the transducer reported in Ref [66]. (b) Photographs of the fabricated flexible transparent CMUT arrays and bond with flexible PCB.[76] (c) Schematic showing the cross-section and ultrasound detection mechanism of a surface-micromachined optical ultrasound transducer element reported in Ref [78]. (d) Schematic diagram and photograph of the fiber optic ultrasound sensor.[81] (e) Design and structure of the sensitive ultrawideband transparent ultrasound detector in Ref [94].

## 6 PA plus Deep Learning

### 6.1 Image Reconstruction and Enhancement

A major focus of deep learning in PAI has been improving image reconstruction, especially under challenging conditions like limited-view geometries, where the acquired data is sparse. Methods like DL-PAT[97] and DuDoUNet[98] use convolutional neural networks to reduce artifacts and improve image quality from limited-view data, while work by T. Wang et al. proposed a learned regularization approach.

DL-PAT[97] implements a conditional generative adversarial network to enhance 3D dynamic volumetric PA computed tomography. By training on a subset of the transducer elements from conventional systems, DL-PAT can reconstruct high-quality images comparable to full-view methods while using fewer elements. This reduces the cost and data size of PAI systems. Quantitative studies showed DL-PAT reduced artifacts and improved signal-to-noise ratio. D. Seong et al.[99] also introduced deep learning techniques into 3D PA imaging. DuDoUNet[98] is specifically designed for limited view PAT. It uses a U-Net architecture that takes both time-domain and frequency-domain representations of the limited view data as input. This provides complementary information to distinguish artifacts from true signals. An information sharing block fuses and compares the dual-domain inputs. Experiments on a clinical database showed DuDoUNet reconstructed images with 93.56% structural similarity and 20.89 dB peak signal-to-noise ratio, outperforming conventional limited view methods.

Other works have applied deep learning for computational acceleration. For example, the proposed method in Ref[100] achieves faster convergence by learning regularization features. It uses a CNN within a model-based gradient descent reconstruction to learn regularization parameters automatically instead of manual adjustment. AS-Net[101] fuses multi-feature information to enable faster reconstruction from sparse data. It was demonstrated to provide superior image quality from limited data compared to conventional model-based reconstruction.

Beyond image reconstruction, deep learning has been used for resolution enhancement. Works like AR to OR domain transfer learning (AODTL)-GAN[102] train generative adversarial networks (GANs) to transform acoustic-resolution images to optical-resolution quality. AODTL-GAN uses a two-stage GAN approach. First, a generative model is trained on simulated acoustic and optical resolution image pairs. Then, a second network adapts the output to match real optical resolution images through domain transfer learning. Quantitative metrics like peak signal-to-noise ratio and structural similarity index were also significantly increased. S. Cheng et al. employs a GAN to enhance the imaging lateral resolution of acoustic-resolution PAM (AR-PAM) images, transforming them to achieve OR-PAM quality[103]. This GAN enables deep-tissue imaging and demonstrating potential applications in biomedicine.

Others like Deep-E[104] focuses on enhancing elevation resolution in linear-array based systems, which is inherently limited by the transducer geometry. Based on U-Net, Deep-E enhances the elevational resolution of linear-array-based photoacoustic tomography by training on 2D slices in axial and elevational planes, leading to improved resolution by at least four times and potential high-speed, high-resolution image enhancement applications. J. Kim et al.[105] introduces a computational strategy utilizing deep neural networks (DNNs) for enhancing both temporal and spatial resolutions in localization photoacoustic imaging, which is illustrated in Fig. 7(a). The proposed DNN-based method reconstructs high-density supper resolution images from a reduced number of raw frames. This approach is applicable to both 3D label-free localization OR-PAM and 2D labeled localization PAT. C. Dehner et al.[106] proposed DL-MSOT to apply deep learning denoising to improve optoacoustic contrast in deep tissues. The study introduces a deep learning approach for noise removal before image reconstruction. This algorithm learns spatiotemporal noise-signal correlations using entire optoacoustic sinograms and is trained on real noise and synthetic optoacoustic data. Evaluations showed it achieved substantially higher vessel contrast at depths over 2 cm in vivo. Other super-resolution methods include work by Y. Ma et al.[107] and work by D. He et al.[108]

High fidelity deconvolution methods, such as using RRDBNet,[109] also leverage deep learning for resolution improvement. RRDBNet is a deep residual network tailored for image deconvolution. By training on simulated vessels, it showed accurate recovery of features ranging from 30 μm to 120 μm. It also outperformed conventional methods like Richardson-Lucy and model-based deconvolution in recovering multiscale features in phantom and in vivo images. Zhengyuan Zhang et al.[110] discussed the application of photoacoustic imaging for monitoring cancer therapy . By tracking changes in vasculature and oxygenation, the technique provides valuable insights into treatment efficacy. The study demonstrates the potential of photoacoustic imaging as a non-invasive tool for assessing therapeutic response and guiding cancer treatments.

In addition, deep learning methods have been mixed with more physics aspects of photoacoustic imaging, such as fluence compensation[111] and beamforming corrections[112]. Deep learning has also been applied in specialized modalities like improving Bessel-beam performance[113] and processing endoscopic images[114].

A. Madasamy et al.[111] uses Fully Dense U-Net based deconvolution for optical fluence compensation in 3D optoacoustic tomography. Training on heterogeneous breast phantoms showed the method highlighted deep structures with higher contrast compared to reconstruction without fluence correction. The method proposed by S. Jeon et al.[112] applies deep learning for mitigation of speed-of-sound (SoS) aberrations in vivo. This method simultaneously mitigates SoS aberrations, removes streak artifacts, and enhances temporal resolution in both structural and functional in vivo PA images of healthy human limbs and melanoma patients.

For emerging modalities, Y. Zhou et al.[113] combines Bessel beam excitation with deep learning to enhance quantitative performance in multi-parametric photoacoustic microscopy using a conditional generative adversarial network (cGAN). It enables simultaneous high-resolution quantification of hemoglobin metrics and cerebral blood flow in live mouse brains.

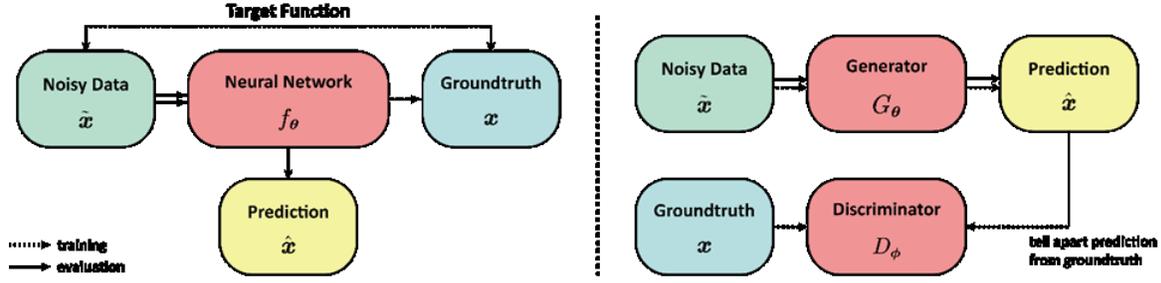

**Fig. 6** A brief comparison of end-to-end neural network methods (Left) and generative adversarial network (GAN) methods (Right). End-to-end methods rely on a well-designed neural network to learn the forward mapping defined by the paired dataset, while GAN needs a generator and a discriminator to perform adversarial learning.

*6.2 Motion Correction and Denoising*

PAI can be affected by motion artifacts and noise. Deep learning approaches have shown promise in addressing these challenges. For example, Motion Artifact Correction (MAC)-Net[115] corrects motion artifacts in intravascular PA data by learning correlations from simulated training data. It uses a convolutional network to correct motion frame-by-frame while preserving structures. Evaluations showed it achieved motion suppression comparable to gating but without discarding frames.

Other works like PA-GAN[116] and the method in Ref[108] train generative adversarial networks to reduce noise and improve signal-to-noise ratio. PA-GAN[116] uses an unpaired training approach that does not require matched noisy and clean image pairs. This provides greater flexibility than supervised methods. Experiments showed PA-GAN achieved higher peak signal-to-noise ratio and structural similarity than U-Net, especially for sparse-view cases. The method in Ref[108] trains a GAN to emulate the effects of both temporal averaging and singular value decomposition (SVD) denoising. It effectively enhances signal-to-noise ratio (SNR) in radio-frequency (RF) data and corresponding photoacoustic reconstructions, leading to reduced scan time and laser dose.

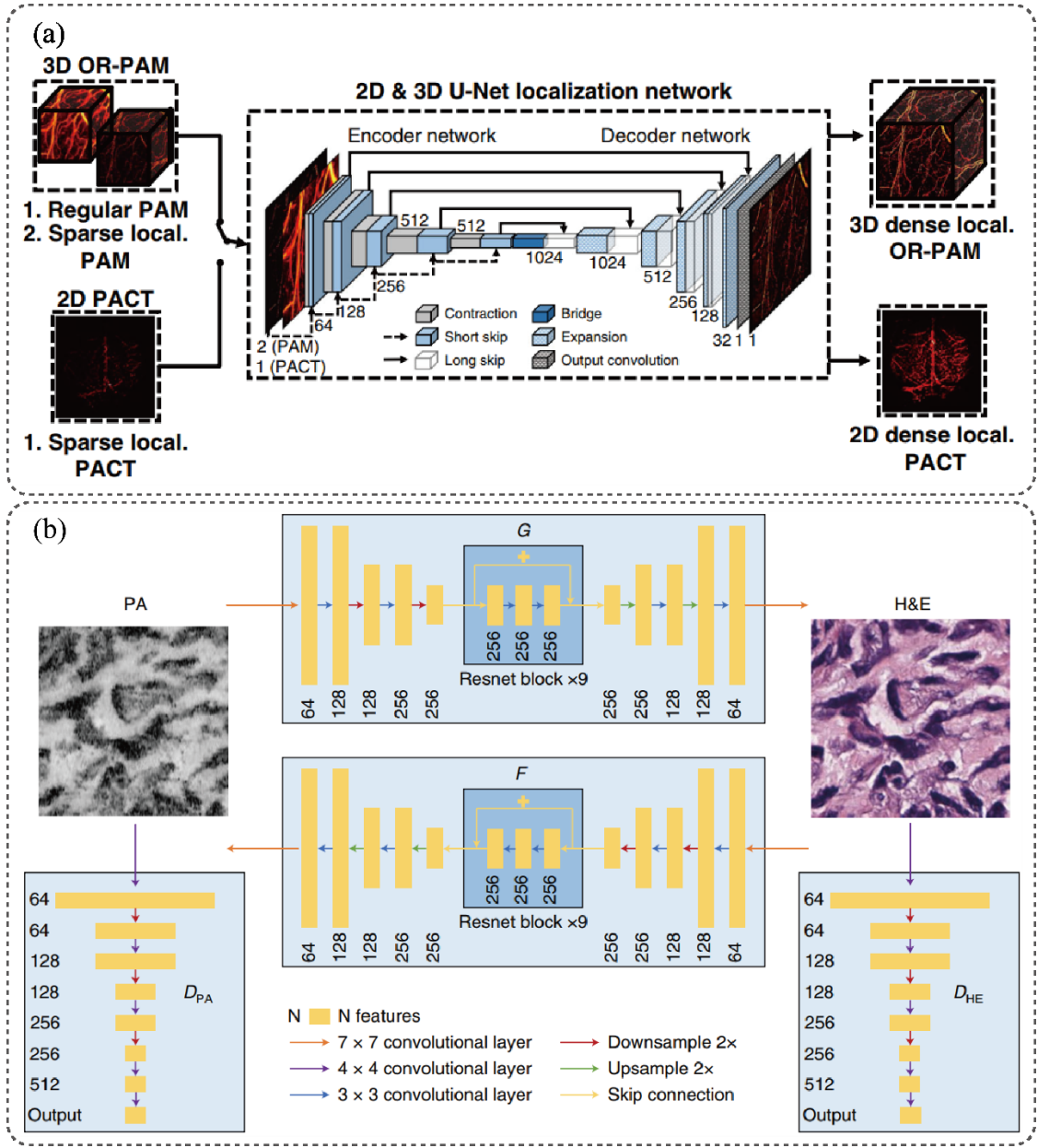

**Fig. 7** Demonstration of two deep learning methods in photoacoustic imaging. (a), 3D U-Net proposed by Jongbeom Kim et al. for reconstruction of high-density superresolution images from fewer raw frames[10]; (b), the structure of the CycleGAN proposed by Rui Cao et al., where the CycleGAN is composed of two symmetric generators and corresponding adversarial discriminators[32].

## 6.3 Analysis and Quantification

Deep learning has also enabled accurate analysis of PA data. Semantic segmentation algorithms can extract interpretable information from multispectral PA images.[117] This study introduces a deep learning-based method for semantic segmentation of multispectral photoacoustic images, using manually annotated data to train a supervised algorithm.

For quantification, methods like quantitative optoacoustic tomography (QOAT)-Net[118] estimate optical absorption coefficients by training on simulated data-label pairs. QOAT-Net uses a dual-path network design that learns correlations between the imaging data and absorption maps to perform fluence compensation. QOAT-Net is demonstrated to produce high-resolution quantitative absorption images in simulations, phantoms, ex vivo, and in vivo tissues. This innovation facilitates DL-based QOAT and similar imaging applications even when ground truth data is unavailable.

US-UNet[119] learns from clinical ultrasound and photoacoustic features for diagnosis. By using ultrasound morphology features from a pretrained CNN along with PA data, it achieved an area under the ROC curve (AUC) of 0.94 and accuracy of 0.89 in differentiating ovarian lesions in patient studies. This demonstrates deep learning can leverage multimodal data for enhanced quantification. H. Zhao et al.[120] proposed a deep learning-based technique for OR-PAM to effectively image and analyze 3D microvasculature datasets. This method overcomes limitations in depth of focus and signal-to-noise ratio, showcasing successful segmentation of endogenous and exogenous multi-organ data. Notably, it achieves comprehensive exogenous 3D imaging of mouse brain vasculature at various depths, highlighting its potential for microcirculation imaging in clinical applications.

*6.4 Other Applications*

Deep learning has opened possibilities for new photoacoustic imaging capabilities. As examples, Melanie Schellenberg et al. proposed a GAN to synthesize realistic tissue images for quantitative PAT.[121] It uses GANs trained on annotated medical images to generate virtual tissue structures. Adding simulated optical and acoustic properties then yields realistic training data. The method is validated against a traditional model-based approach, demonstrating more realistic synthetic images.

Deep learning algorithms have also enabled new modalities like Deep-PAM[122] for rapid label-free histological imaging. Deep-PAM combines ultraviolet PAM (UV-PAM) with deep learning to enable rapid and label-free histological imaging. This provides rapid label-free assessment of specimens without physical staining or tissue processing. Besides, Cao et al.[123] introduces a real-time, label-free method for intraoperative evaluation of thick bone specimens using UV-PAM in reflection mode, as is shown in Fig. 7 (b). This technique eliminates the need for tissue sectioning and provides detailed three-dimensional contour scans of bone tissue.

## 7 PA plus Other Imaging Modality

### 7.1 Enabling Multimodal Medical Imaging

Breast imaging is one of the clinical applications of PA imaging, and there have been numerous studies on breast PA imaging[124], including the development of PA-US dual-modal systems. Zheng et al. incorporated ultrasound elastography as a modality in addition to PA and US imaging to assess the mechanical properties of the breast[125, 126]. The system employed motor control and a 3D-printed transducer-fiber combiner to reduce registration errors, enabling a 3D scan of the breast to be completed in approximately 40 seconds.

Brain imaging is another emerging application of PA imaging, holding great promise for studying brain functionality[127-129]. Na et al.[130] combined PA imaging with functional ultrasound imaging in the CRUST-PAT system, which employed cross-line ultrasound tomography to provide all-directional sensitivity to blood flow, enabling simultaneous monitoring of cerebral blood flow and oxygenation. The spatial resolution achieved was around 170 micrometers. The imaging results are shown in Fig. 8 (a)-(f). Ni et.al.[131] successfully demonstrated the imaging of the superior middle cerebral vein in the temporal cortex of a healthy volunteer using multi-spectral optoacoustic tomography (MSOT) and time-of-flight magnetic resonance angiography (MRA). These initial results show the potential of MSOT in clinical brain imaging. However, the human skull induces strong acoustic aberrations, resulting in significant distortion of deep vascular structures. In these studies, the presence of the skull had an impact on the signals, leading to a reduction in image resolution. Therefore, PA transcranial imaging remains a major challenge.

There are also numerous emerging applications of PA imaging that have been developed.[132] Chen et al.[133] applied a PA-US dual-modal imaging system for assessing the activity of scar tissue. The current methods for evaluating scars primarily rely on subjective assessments by physicians. This team utilized PA imaging, ultrasound imaging, elastography, and super micro vascular imaging to evaluate scar tumors and performed standardized quantitative assessments. Clinical experiments also demonstrated the feasibility of this evaluation model. Wang et al. [134] used a PAUS dual-modal imaging system for diagnosing rheumatoid arthritis, a condition characterized by neovascularization, synovial hyperplasia, and cartilage damage. The team employed PA imaging to detect blood vessel formation and employed US to assess synovial erosion, correlating them with the severity of arthritis for quantitative analysis.

*7.2 Exciting PA Signal for Ultrasound Imaging*

Combining PA and US techniques can overcome some limitations of traditional ultrasound imaging while utilizing PA to assist in ultrasound generation. Typically, a fully optical US probe uses two separate fibers for ultrasound generation and reception respectively. However, Chen et al.[135] incorporated PA-based US generation and Fabry-Perot (FP) interference-based ultrasound detection structures at the end of a double-clad optical fiber. This innovative approach allowed them to maintain a probe size of just 1 mm in diameter. Experimental results demonstrated that the probe was capable of producing ultrasound signals with an amplitude of 2.36 MPa, central frequency of 10.64 MHz, and a -6-dB bandwidth of 22.93 MHz in transmitting mode. The researchers also successfully captured forward-viewing pulse-echo signals that varied with transmission distance for the first time.

In another study, Liu et al.[136] developed a fiber-optic ultrasound pulse transmitter based on continuous-wave (CW) laser-triggered thermo-cavitation. By heating a highly absorptive copper nitrate solution using CW laser light, they generated explosive bubbles and emitted strong ultrasound waves. Operating at a wavelength of 980 nm and with an optical heating power of 50 mW, they achieved omnidirectional ultrasound pulses with an intensity of 0.3 MPa, a repetition rate of 5 kHz, within the frequency range of 5-12 MHz. They used this ultrasound transmitter to construct an all-fiber ultrasound endoscopic probe, eliminating the need for expensive high-energy pulsed lasers and optically absorptive composite films.

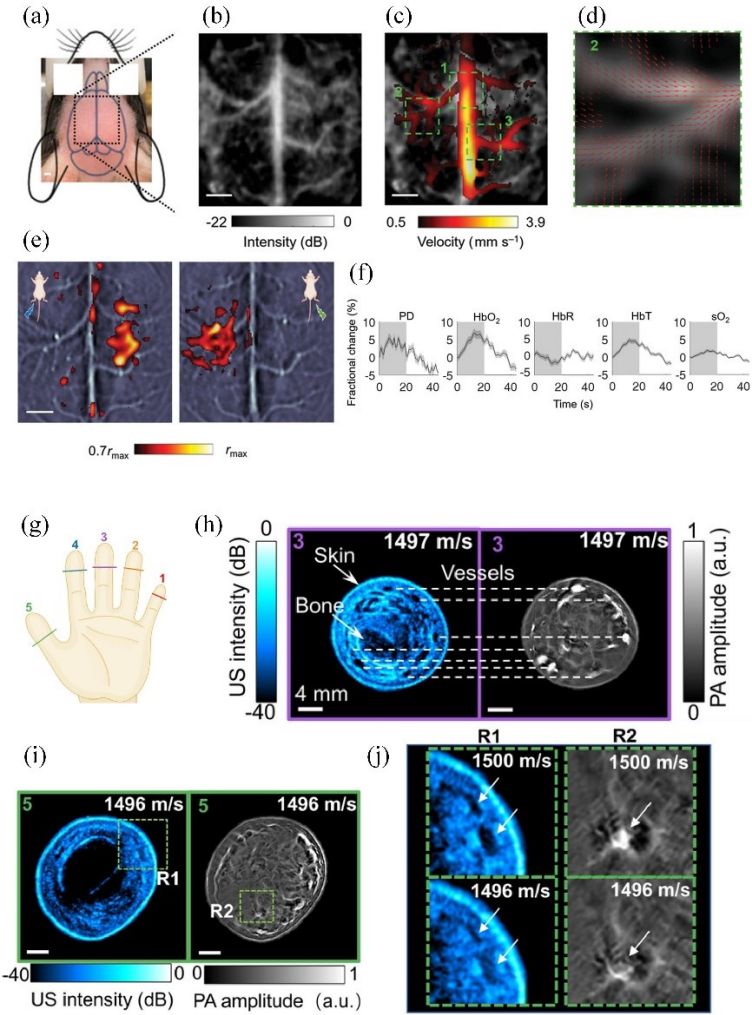

**Fig. 8** (a)-(f) demonstrate the results of transcranial brain imaging by Na et al.[130] (g)-(j) demonstrates the imaging result of human finger joints with VF-USPACT by Zhang et al.[137] (a) The regions imaged in transcranial brain imaging of mice. (b) power Doppler images(PDI) of the mouse brain. (c) Velocity amplitude map of the cerebral blood flow (CBF) in the major cortical vessels. (d) Flow vectors in regions 2 of the velocity map. (e) PAT-measured functional responses, which show the contralateral functional responses to the hindlimb electrical stimulation. (f) Fractional changes of PD, hemoglobin concentrations, and sO2 signal in response to stimulation. (g) Illustration of the finger joint imaging locations. (h) PA and US image reconstructed with optimal SoS at middle finger cross section. The high PA signals from blood vessels corresponding to anechoic regions in the US images are shown. (i) PA and US image of the thumb cross section. (j) The comparison of the image reconstructed with optimized SoS (1496m/s) and the one from little finger (1500m/s).

*7.3 Improving Algorithm Design*

The presence of US can assist in optimizing the reconstruction algorithm in PA imaging.[138] In PA imaging of practical scenarios, the presence of heterogeneous media could cause the artifacts and lead to decreased image quality. One approach is to use physics-based iterative optimization algorithms, which is, however, time-consuming[139]. By combining US imaging, which provides structural information, it becomes possible to provide prior information about the sound velocity field during PA image reconstruction[140, 141].

Zhang et al.[142] employed time-of-flight (TOF) for automatic segmentation of sample boundaries and automatically searched for the optimal sound speed, demonstrating good robustness. However, their search process was relatively time-consuming, and the image reconstruction took approximately 14 minutes. They further[137] developed a real-time, 10 Hz, dual-modal system combining ultrasound and PA imaging. The system leveraged ultrasound imaging to automatically determine the optimal sound speed and employed selective parallel image reconstruction techniques to enhance the imaging speed. The imaging results of human fingers are shown in Fig. 8 (g)-(j).

To address the presence of bone, Zhao et al.[140] utilized US imaging to identify the acoustically heterogeneous regions within the measurement area and segmented those regions. Subsequently, they applied a variable truncation time to truncate the PA signals in the time domain, effectively suppressing acoustic artifacts. However, this method is specifically designed for imaging outside the high-speed regions and may not be suitable for scenarios such as transcranial imaging which requires the penetration through high-density areas.

*7.4 Multimodal Endoscopic Imaging*

PA endoscopy (PAE), as a clinical application of PAI, has been under development for a significant period. However, there are still great obstacles to overcome before its clinical implementation can be achieved. To address this concern, the following studies underscore the significant progress made to promote the progress of PAE toward clinical application[143-147].

Major difficulty in multimodal PAE lies in achieving efficient light coupling. Wen et al.[148] addressed this issue by introducing a disposable PAUS endoscopic catheter prototype and its corresponding power interface unit. The catheter exhibited switchability, self-internal 3D scanning, and system repeatability for gastrointestinal endoscopy. By utilizing high-fluence relays, they minimized the cascade insertion losses of the optical waveguide to 0.6 dB while maintaining high-power impedance performance. They also designed a customized focus-adjustable acousto-optic coaxial probe for high-sensitivity optical-resolution PA imaging. Their experiments demonstrated real-time microscopic visualization of microvasculature and stratification in the rat colon, with a lateral resolution of 18 μm and an axial resolution of 63 μm. The rigid part of the catheter had a length of 13 mm, showcasing significant potential for clinical gastrointestinal disease detection.

In another study, Zhu et al.[149] addressed the environmental requirements of endoscopy by integrating a miniaturized ultrasound array and an angle-tipped optical fiber into a hydrostatic balloon catheter. The flexible surface of the hydrostatic balloon enabled acoustic coupling on the uneven surfaces of the gas-filled intestine. This integration of endoscopic PA imaging technology allowed for the evaluation of colitis and fibrosis feasibility. When the balloon was collapsed, the catheter probe could potentially be compatible with clinical ileo-colonoscopy. With an imaging penetration depth of 12 mm, they validated the probe's potential in differentiating normal, acute, and chronic conditions in intestinal obstruction using an in vivo rabbit model.

## 8. Outlook and Conclusion

In this review, we summarized several typical topics of PAI plus X innovations. More generally, PAI plus X strategy can be classified as two categories: PAI empowered by X, and PAI employing X. Simply speaking, PAI empowered by X means PAI is made better by X. On the other hand, PAI employing X means PAI makes X better. As shown in Fig. 9, we summarize the outlook and challenges of PAI plus X in a wider scope, in such two categories.

For PAI empowered by X, it can be further classified as: new hardware and new algorithms. Within new hardware, PAI can be empowered by new laser sources, acoustic sensors, mechanical components, and electrical circuits. Within new algorithm, PAI can be empowered by various algorithm designs. Each of these hardware and algorithms improves the overall performance of the PAI system, meanwhile there is still room for further improvement in many aspects (rightmost column in Fig. 9). Just give an example, for learning-based PAI reconstruction algorithm, large clinical dataset is still a bottleneck, impeding the deep learning algorithms for clinical applications. Similarly, for PAI empowering X, it can be further classified as: treatment guidance, ultrasound generation and multimodal imaging. Within treatment guidance, PAI can guide various treatment methods, such as photothermal/HIFU/RF therapy, as well as surgical robot. Within ultrasound generation, PA generated ultrasound signal can be used for neuro-stimulation and ultrasound imaging. Within multimodal imaging, PAI can provide functional and molecular information, complementing the anatomical imaging of conventional imaging modality, such as B-mode ultrasound and CT. Some key parameters and advantages are listed in Fig. 9.

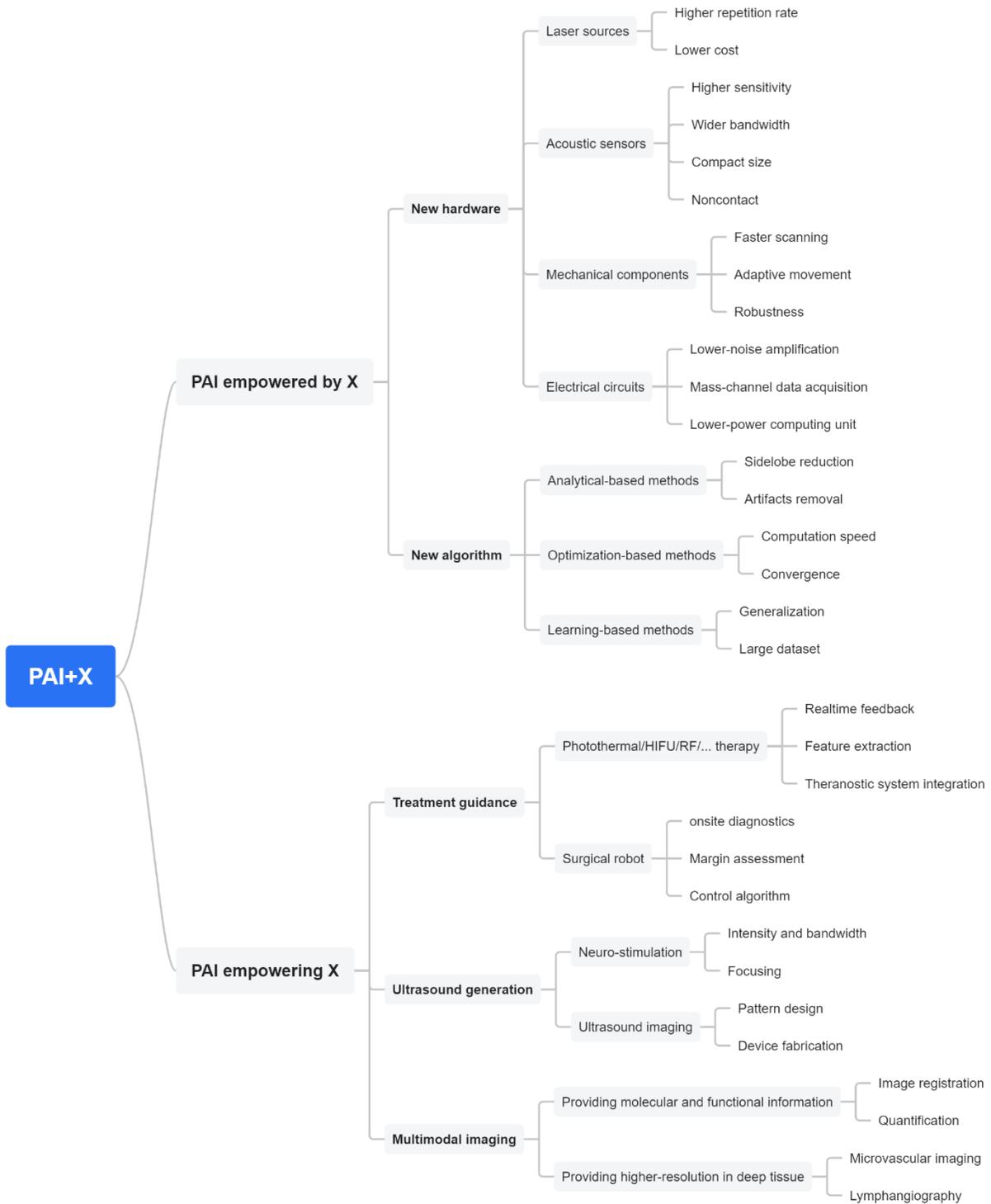

**Fig. 9** Outlook of PAI plus X.

**Acknowledgement**

This research was funded by National Natural Science Foundation of China (61805139), Shanghai Clinical Research and Trial Center (2022A0305-418-02), and Double First-Class Initiative Fund of ShanghaiTech University (2022X0203-904-04). There is no relevant financial interests in this manuscript.